\documentclass[showpacs,aps,prd,superscriptaddress,nofootinbib,floatfix,amsmath,amssymb]{revtex4}
\usepackage{graphicx}
\usepackage{amssymb}

\begin{document}


\title{Static quantities of the $W$ boson in the ${SU_L(3)}\times {U_X(1)}$
 model with right-handed neutrinos}
\author{J. L. Garc\'\i a-Luna}
\email[E-mail:]{jluna@cucei.udg.mx}\affiliation{Departamento de
F\'{\i}sica,  Centro Universitario de Ciencias Exactas e
Ingenier\'{\i}as, Universidad de Guadalajara, Blvd. Marcelino
Garc\'{\i}a Barrag\'an 1508, C.P. 44840, Guadalajara Jal.,
M\'exico}
\author{G. Tavares-Velasco}
\email[E-mail:]{gtv@fcfm.buap.mx} \affiliation{Facultad de
Ciencias F\'\i sico Matem\' aticas, Benem\' erita Universidad
Aut\' onoma de Puebla, Apartado Postal 1152, Puebla, Pue.,
M\'exico}
\author{J. J. Toscano}
\email[E-mail:]{jtoscano@fcfm.buap.mx} \affiliation{Facultad de
Ciencias F\'\i sico Matem\' aticas, Benem\' erita Universidad
Aut\' onoma de Puebla, Apartado Postal 1152, Puebla, Pue.,
M\'exico}

\date{\today}

\begin{abstract}
The static electromagnetic properties of the $W$ boson, $\Delta
\kappa$ and $\Delta Q$, are calculated in the ${SU_L(3)}\times
{U_X(1)}$ model with right-handed neutrinos. The new contributions
from this model arise from the gauge and scalar sectors. In the
gauge sector there is a new contribution from a complex neutral
gauge boson $Y^0$ and a singly-charged gauge boson $Y^\pm$. The
mass of these gauge bosons, called bileptons, is expected to be in
the range of a few hundreds of GeV according to the current bounds
from experimental data. If the bilepton masses are of the order of
200 GeV, the size of their contribution is similar to that
obtained in other weakly coupled theories. However the
contributions to both $\Delta Q$ and $\Delta \kappa$ are
negligible for very heavy or degenerate bileptons. As for the
scalar sector, an scenario is examined in which the contribution
to the $W$ form factors is identical to that of a
two-Higgs-doublet model. It is found that this sector would not
give large corrections to $\Delta \kappa$ and $\Delta Q$.

\end{abstract}


\maketitle

\section{Introduction}
\label{introduction} The experimental scrutiny of the Yang-Mills
sector is essential to test the standard model (SM). In
particular, the trilinear gauge boson couplings (TGBCs) offer a
unique opportunity to find evidences of new physics through the
study of their one-loop corrections. Hopefully, the TNGBCs will be
tested with unprecedent accuracy beyond the tree level at hadronic
and leptonic colliders in the near future \cite{Wudka}. Particular
emphasis has been given to the study of the static quantities of
the $W$ boson. The CP-even electromagnetic properties of the $W$
boson are characterized by two form factors, $\Delta \kappa$ and
$\Delta Q$, which are the coefficients of Lorentz tensors of
canonical dimension 4 and 6, respectively \cite{Hagiwara}. Both
form factors can only arise at the one-loop level within the SM
and other renormalizable theories, thereby being sensitive to
sizeable new physics effects. It has been argued that $\Delta Q$
is not sensitive to nondecoupling effects and thus it could only
be useful to search for effects of new physics near the Fermi
scale \cite{Inami}. On the contrary $\Delta \kappa$ might be
sensitive to heavy physics effects due to its nondecoupling
properties \cite{Inami}. The one-loop contributions to $\Delta Q$
and $\Delta \kappa$ were long ago studied in the SM
\cite{Bardeen,Argyres} and more recently in the context of several
specific theories \cite{THDM,WWg-NP,Tavares}. Also, a
model-independent study of the $WW\gamma$ vertex via the effective
Lagrangian approach was presented in \cite{Papavassiliou}.

In a recent work \cite{Tavares} two of us studied the static
quantities of the $W$ boson in the context of the minimal
${SU_L(3)}\times {U_X(1)}$ model, dubbed as 331 model
\cite{Pleitez1}. The main attraction of this model is the unique
mechanism of anomaly cancellation, which is achieved provided that
all of the fermion families are summed over rather than within
each fermion family, as occurs in the SM. As a consequence, the
number of fermion families must be a multiple of the quark color
number, which offers a possible solution to the flavor problem.
The 331 model has been the source of considerable interest
recently \cite{Tavares-331}. In this work we will focus on the
contributions to $\Delta \kappa$ and $\Delta Q$ from both the
gauge and scalar sectors of the 331 model with right-handed
neutrinos \cite{Long}. This version is attractive because, in
order to achieve the mechanism of spontaneous symmetry breaking
(SSB) and generate the gauge boson and fermion masses, it requires
a Higgs sector which is more economic than that of the minimal
version \cite{Long}. Evidently the features of the 331 model with
right-handed neutrinos are rather different than those of the
minimal version, and so are the contributions to the static
quantities of the $W$ boson. It is thus worth evaluating the
behavior of $\Delta \kappa$ and $\Delta Q$ in the new scenario
raised by this model.  Special attention will be paid to discuss
the contribution arising from the gauge sector because it is the
one which has the more interesting features. As will be shown
below, the contribution from the scalar sector is similar to that
arising in a two-Higgs-doublet model (THDM) \cite{THDM}.

A peculiarity of 331 models is that they predict a pair of massive
gauge bosons arranged in a doublet of the electroweak group, which
emerge when ${SU_L(3)}\times {U_X(1)}$ is broken into
${SU_L(2)}\times {U_Y(1)}$. While the minimal 331 model predicts a
pair of singly-charged and a pair of doubly-charged gauge bosons,
the model with right-handed neutrinos predicts a pair of neutral
no self-conjugate gauge bosons $Y^{0}$ instead of the
doubly-charged ones. These new gauge bosons are called bileptons
since they carry lepton number $L =\pm \,2$, and thus are
responsible for lepton-number violating interactions
\cite{Cuypers}. It has been pointed out that the neutral bilepton
is a promising candidate in accelerator experiments since it may
be the source of neutrino oscillations \cite{Long-Inami}. Very
interestingly, the couplings between the SM gauge bosons and the
bileptons do not involve any mixing angle and are similar in
strength to the couplings existing between the SM gauge bosons
themselves. Current bounds establish that the bilepton  masses may
be in the range of a few hundred of GeVs \cite{Bounds}. It is then
feasible that these bileptons may show up through their virtual
effects in low-energy processes. This is an important reason to
investigate the effect of these particles on the $WW\gamma$ vertex
function. It is also interesting that, due to the SSB hierarchy,
the splitting between the bilepton masses $m_{Y\pm}$ and $m_{Y^0}$
is bounded by $m_{W}$, so the bileptons would be almost degenerate
since their masses are expected to be heavier than the $W$ mass.
Therefore, the gauge boson contribution to the static quantities
of the $W$ boson would depend on only one free parameter. As far
as the scalar sector is concerned, this model predicts the
existence of ten physical scalar bosons \cite{LongHP}: four
neutral CP-even, two neutral CP-odd and four charged ones. From
these scalar bosons, only three of them, two neutral CP-even and a
charged one, couple with the $W$ boson at the tree level because
they are the only ones which get their masses at the Fermi scale.
Furthermore, in order to reproduce the SM at low energies, we will
concentrate on a scenario in which one of the neutral Higgs bosons
coincides with the SM Higgs boson.

The rest of the paper is organized as follows. Sec. \ref{model} is
devoted to a brief description of the $331$ model with
right-handed neutrinos. In Sec. \ref{calculation} we present the
calculation of the static properties of the $W$ boson. The
numerical results are analyzed in Sec. \ref{discussion}, and the
conclusions are presented in Sec. \ref{conclusions}.

\section{Brief review of the 331 model with right-handed neutrinos}
\label{model} The features of the 331 model with right-handed
neutrinos has been discussed to a large extent in Ref.
\cite{Long}. Here we will only review those aspects which are
relevant for the present discussion. The fermion spectrum of the
model is shown in Table \ref{spectrum}. The three lepton families
are arranged in triplets of $SU_L(3)$, whereas in the quark sector
it is necessary to introduce three exotic quarks $D_1$, $D_2$, and
$T$. In order to cancel the $SU_L(3)$ anomaly, two quark families
must transform as $SU_L(3)$ antitriplets and the remaining one as
a triplet. It is customary to arrange the first two quark families
in antitriplets and the third one in a triplet. This choice is
meant to distinguish the possible new dynamics effects arising in
the third family.

\begin{table}[!htb]
\begin{tabular}{ccc}
\hline \hline
Leptons&First two quark families&Third quark family\\
\hline \hline $
f^i_L=\left( \begin{array}{ccc} \nu^i_L \\
e^i_L\\
(\nu^c_L)^i
\end{array}\right) \sim (1,3,-1/3)$
&$Q^{i}_L=\left( \begin{array}{ccc} d^{i}_L \\
-u^{i}_L\\
D^{i}_L
\end{array}\right) \sim (3,\bar{3},0)$&
$Q^3_L=\left( \begin{array}{ccc} u^3_L \\
d^3_L\\
T_L
\end{array}\right) \sim (3,3,1/3)$\\
$e^i_R \sim (1,1,-1)$&$\begin{array}{lll}
u^{i}_R \sim (3,1,2/3)\\
d^{i}_R \sim (3,1,-1/3)\\
D^{i}_R \sim(3,1,-1/3)
\end{array}$&
$\begin{array}{lll}
u^3_R\sim (3,1,2/3)\\
d^3_R\sim (3,1,-1/3)\\
T_R\sim (3,1,2/3)
\end{array}$\\ \hline \hline
\end{tabular}
\caption{\label{spectrum}Fermion spectrum of the 331 model with
right-handed neutrinos, along with the quantum number
assignments.}
\end{table}

The electric charges of the exotic quarks are $Q_{D_1\,D_2}=-1/3\,
e$ and $Q_T=2/3\, e$. This is to be contrasted with the three new
quarks, $D$, $S$ and $T$, predicted by the minimal 331 model,
whose charge is indeed exotic, namely $Q_{D,\,S}=-4/3\,e$ and
$Q_T=5/3\,e$.

As already mentioned, the 331 model with right-handed neutrinos
has the advantage that it requires a Higgs sector simpler than
that introduced in the minimal version. In fact, only three
triplets of $SU_L(3)$ are needed to reproduce the known physics at
the Fermi scale:

\begin{eqnarray}
\chi=\left( \begin{array}{cc} \Phi_3 \\
 \chi^{'0}
\end{array}\right)\sim (1,3,-1/3),\ \
\rho=\left( \begin{array}{cc} \Phi_1 \\
 \rho^{'+}
\end{array}\right)\sim (1,3,2/3),\ \
\eta=\left( \begin{array}{ccc} \Phi_2 \\
 \eta^{'0}
\end{array}\right)\sim (1,3,-1/3),
\end{eqnarray}
where $\Phi^\dag_1=(\rho^-,\rho^{0*})$,
$\Phi^\dag_2=(\eta^{0*},\eta^+)$, and
$\Phi^\dag_3=(\chi^{'0*},\chi^+)$ are $SU_L(2)\times U_Y(1)$
doublets with hypercharge $+1$, $-1$, and $-1$, respectively. This
is to be contrasted again with the minimal 331 model, which
requires the presence of three triplets and one sextet
\cite{Pleitez1}. The vacuum expectation value (VEV)
$<\chi>^T=(0,0,w/\sqrt{2})$ breaks down the $SU_L(3)\times U_N(1)$
group into $SU_L(2)\times U_Y(1)$. In this first stage of SSB, the
new gauge bosons and quarks, as well as some physical scalars, get
their masses. At the Fermi scale all the known SM particles and
some physical scalar bosons are endowed with masses through the
VEV $<\Phi_1>=(0,v_1/\sqrt{2})$ and $<\Phi_1>=(0,v_1/\sqrt{2})$.
In this way, the $\Phi_1$ and $\Phi_2$ doublets break the
$SU_L(2)\times U_Y(1)$ group into $U_e(1)$.

In addition to the three exotic quarks, the model predicts the
existence of five new gauge bosons: two singly charged $Y^\pm$,
two neutral no self-conjugate $Y^0$, and a neutral self-conjugate
$Z'$. The $Y^\pm$ and $Y^0$ gauge bosons are called bileptons
because they carry two units of lepton number \cite{Cuypers}. All
the new particles acquire their masses at the $w$ scale. At this
stage of SSB, the exotic quarks together with the $Z'$ boson
emerge as singlets of the electroweak group. Consequently, these
particles cannot interact with the $W$ boson at the tree level. It
follows that there is no contribution from the exotic quarks to
the static electromagnetic properties of the $W$ boson at the
lowest order. As for the $Z'$ boson, it couples to the $W$ boson
via the $Z'-Z$ mixing induced at the Fermi scale, which means that
the respective contribution to the $WW\gamma$ vertex is expected
to be strongly suppressed since it is proportional to the
corresponding mixing angle, which is expected to be negligibly
small \cite{Long}. On the other hand, the dynamical behavior of
the bileptons is different since they arise as a doublet of the
electroweak group at the $w$ scale and thus have nontrivial
couplings with the SM gauge bosons. Due to the fact that the
$SU_L(2)$ group is completely embedded in $SU_L(3)$, the bileptons
couple with the SM gauge bosons with a strength similar to that of
the couplings existing between the SM gauge bosons. In particular,
these new couplings are entirely determined by the coupling
constant associated with $SU_L(2)$ and the weak angle $\theta_W$.
When $SU_L(2)\times U_Y(1)$ is broken down to $U_e(1)$, the masses
of the bileptons receive new contributions. In the gauge sector,
the mass eigenstates arise from the Higgs kinetic-energy term,
which is given by
\begin{equation}
\mathcal{L}=(D_\mu\chi)^\dag(D^\mu \chi)+(D_\mu \rho)^\dag(D^\mu
\rho)+(D_\mu\eta)^\dag(D^\mu \eta),
\end{equation}
where $D_\mu$ is the covariant derivative associated with the
$SU_L(3)\times U_N(1)$ group, which in the fundamental
representation is given by
\begin{equation}
D_\mu=\partial_\mu
-g\frac{\lambda^a}{2}A^a_\mu-ig_N\frac{\lambda^9}{2}N_\mu,
\end{equation}
with $\lambda^a$ $(a=1,2,\cdots ,8)$ being the Gell-man matrices
and $\lambda^9=\sqrt{2/3}\,{\rm diag}(1,1,1)$. Once this sector is
diagonalized, there emerge the following mass-eigenstate fields:

\begin{eqnarray}
&&Y^0_\mu=\frac{1}{\sqrt{2}}\left(A^4_\mu-iA^5_\mu\right), \\
&&Y^-_\mu=\frac{1}{\sqrt{2}}\left(A^6_\mu-iA^7_\mu\right),\\
&&W^+_\mu=\frac{1}{\sqrt{2}}\left(A^1_\mu-iA^2_\mu\right),
\end{eqnarray}
with masses $m^2_{Y^0}=g^2(w^2+v^2_2)/4$,
$m^2_{Y^\pm}=g^2(w^2+v^2_1)/4$, and $m^2_W=g^2(v^2_1+v^2_2)/4$.
From these expressions, it is easy to see that there is an upper
bound on the splitting between the bilepton masses:
\begin{equation}
\label{splitting} \left|m^2_{Y^0}-m^2_{Y^\pm}\right|\leq m^2_W.
\end{equation}
The remaining three gauge fields $A^3_\mu$, $A^8_\mu$,and $N_\mu$
define the self-conjugate mass eigenstates $A_\mu$, $Z_\mu$, and
$Z'_\mu$ \cite{Long}. As far as the Yang-Mills sector of the model
is concerned, it is given by

\begin{equation}
{\cal L}_{YM}=-\frac{1}{4}F^a_{\mu \nu}F^{\mu
\nu}_a-\frac{1}{2}N_{\mu \nu}N^{\mu \nu},
\end{equation}
where $F^a_{\mu \nu}=\partial_\mu A^a_\nu-\partial_\nu
A^a_\mu+f^{abc}A^b_\mu A^c_\nu$ and $N_{\mu \nu}=\partial_\mu
N_\nu-\partial_\nu N_\mu$, being $f^{abc}$ the structure constants
associated with $SU_L(3)$. After this Lagrangian is expressed in
terms of mass eigenstate fields, it can be split into three
$SU_L(2)\times U_Y(1)$-invariant pieces:
\begin{equation}
\label{L_YM} {\cal L}_{YM}={\cal L}^{SM}_{YM}+{\cal
L}^{SM-NP}_{YM}+{\cal L}^{NP}_{YM},
\end{equation}
where the first term represents the well known Yang-Mills sector
associated with the electroweak group, whereas ${\cal
L}^{SM-NP}_{YM}$ represents the interactions between the SM gauge
fields and the heavy ones:
\begin{eqnarray}
\label{L_SM-NP} {\cal L}^{SM-NP}_{YM}&=&-\frac{1}{2}\left(D_\mu
Y_\nu-D_\nu Y_\mu\right)^\dag \left(D^\mu Y^\nu-D^\nu Y^\mu\right)
-iY^\dag_\mu\left( g{\bf F}^{\mu \nu} +g'{\bf B}^{\mu
\nu}\right)Y_\nu\nonumber \\
&-&\frac{ig}{2}\frac{\sqrt{3-4s^2_W}}{c_W}
Z^\prime_{\mu}\Big(Y^\dag_\nu \left(D^\mu Y^\nu-D^\nu
Y^\mu\right)-\left(D^\mu Y^\nu-D^\nu Y^\mu\right)^\dag Y_\nu
\Big),
\end{eqnarray}
where $Y^\dag_\mu=(Y^{0*}_\mu, Y^+_\mu)$ is a doublet of the
electroweak group with hypercharge $-1$, and
$D_\mu=\partial_\mu-ig{\bf A}_\mu +ig'{\bf B}_\mu$ is the
covariant derivative associated with this group. We have
introduced the definitions ${\bf F}_{\mu \nu}=\sigma^iF^i_{\mu
\nu}/2$, ${\bf A}_\mu=\sigma^iA^i_\mu/2$, and ${\bf
B}_\mu=YB_\mu/2$, with $\sigma^i$ the Pauli matrices. Finally, the
last term in Eq. (\ref{L_YM}) is also invariant under the
electroweak group and comprises the interactions between the heavy
gauge fields. There are no contributions to the $WW\gamma$ vertex
arising from this Lagrangian and we refrain from presenting the
respective expression here.

\begin{figure}[!hbt]
\centering
\includegraphics[height=3.6in]{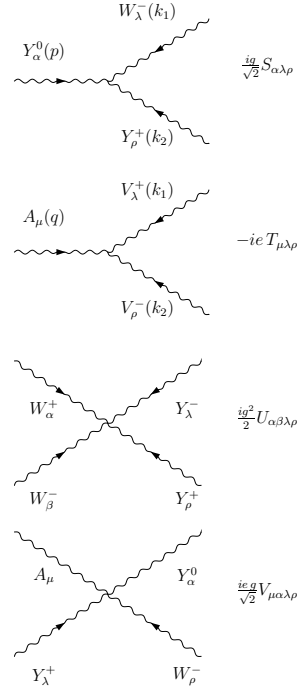}
\caption{\label{FeynRules}Unitary gauge Feynman rules for the
vertices which contribute to the on-shell $WW\gamma$ vertex in the
gauge sector of the 331 model with right-handed neutrinos. The
arrows represent the flow of the 4-momenta.
$S_{\alpha\lambda\rho}=(p-k_2)_\lambda g_{\rho
\alpha}+(k_2-k_1)_\alpha g_{\lambda \rho}+(k_1-p)_\rho g_{\alpha
\lambda}$, $T_{\mu\lambda\rho}= (k_2-k_1)_\mu g_{\lambda
\rho}+(q-k_2)_\lambda g_{\mu \rho} +(k_1-q)_\rho g_{\mu \lambda}$,
$U_{\alpha\beta\lambda\rho}= 2g_{\alpha \rho}g_{\beta
\lambda}-g_{\alpha \beta}g_{\lambda \rho}-g_{\beta \rho}g_{\alpha
\lambda}$, and $V_{\mu\alpha\lambda\rho}= g_{\alpha
\lambda}g_{\rho \mu}-2g_{\alpha \mu}g_{\lambda \rho}+g_{\alpha
\rho}g_{\lambda \mu}$.}
\end{figure}

In the unitary gauge, the contributions to the $WW\gamma$ vertex
arise from the first two terms of the Lagrangian ${\cal
L}^{SM-NP}_{YM}$. These contributions are given by the vertices
$W^\pm Y^\mp Y^0$, $W^\pm W^\mp \gamma$, $Y^\pm W^\mp Y^0 \gamma$,
$W^\pm W^\mp Y^\pm Y^\mp$, and $W^\pm Y^\mp Y^0 \gamma$. The
corresponding Feynman rules are represented in Fig.
\ref{FeynRules}.

As far as the scalar sector is concerned, it was analyzed in
detail in Refs. \cite{Long,LongHP}. Although the most general
Higgs potential is very cumbersome, it gets simplified to a large
extent if one assumes the discrete symmetry $\chi \to -\chi$
\cite{Long,LongHP}. Under this assumption, the scalar potential
can be written in the following way
\begin{eqnarray}
V\left(\chi,\rho,\eta\right)&=&\mu^2_1\left(\eta^\dag
\eta\right)+\mu^2_2\left(\rho^\dag
\rho\right)+\mu^2_3\left(\chi^\dag
\chi\right)+\lambda_1\left(\eta^\dag
\eta\right)^2+\lambda_2\left(\rho^\dag
\rho\right)^2+\lambda_3\left(\chi^\dag
\chi\right)^2\nonumber \\
&+&\left(\eta^\dag \eta\right)\left[\lambda_4\left(\rho^\dag
\rho\right)+\lambda_5\left(\chi^\dag
\chi\right)\right]+\lambda_6\left(\rho^\dag
\rho\right)\left(\chi^\dag
\chi\right)+\lambda_7\left(\rho^\dag \eta\right)\left(\eta^\dag \rho\right)\nonumber \\
&+&\lambda_8\left(\chi^\dag \eta\right)\left(\eta^\dag
\chi\right)+\lambda_9\left(\rho^\dag \chi\right)\left(\chi^\dag
\rho\right)+\lambda_{10}\left(\chi^\dag \eta+\eta^\dag
\chi\right)^2.
\end{eqnarray}
It is worthwhile to analyze the behavior of the scalar potential
at the first stage of SSB. To this end we split
$V\left(\chi,\rho,\eta\right)$ into the following two terms
\begin{equation}
V(\chi,\rho,\eta)=V(\Phi_1,\Phi_2)+V_w,
\end{equation}
with
\begin{equation}
V(\Phi_1,\Phi_2)=\mu^2_1(\Phi^\dag_2
\Phi_2)+\mu^2_2(\Phi^\dag_1\Phi_1)+\lambda_1(\Phi^\dag_2
\Phi_2)^2+\lambda_2(\Phi^\dag_1\Phi_1)^2+\lambda_4(\Phi^\dag_1\Phi_1)(\Phi^\dag_2
\Phi_2)+\lambda_7(\Phi^\dag_1\Phi_2)(\Phi^\dag_2\Phi_1),
\end{equation}
and $V_w$ an intricate function which includes all those terms not
appearing in $V(\Phi_1,\Phi_2)$. $V_w$ is not relevant for the
present discussion, so we will refrain from presenting its
explicit form here. We will content ourselves with mentioning that
this term generates the heavy Higgs boson masses, {\it i.e.} those
which are proportional to the $w$ scale, whereas the the SM gauge
bosons and the remaining physical scalar bosons receive their
masses from $V(\Phi_1,\Phi_2)$ at a relatively light scale. Note
that $V(\Phi_1,\Phi_2)$ corresponds to the scalar potential of a
THDM and so there are five Higgs bosons, which are relatively
light. In fact, the explicit diagonalization of
$V(\chi,\rho,\eta)$ leads to five light and five heavy scalar
bosons \cite{LongHP}. The light scalar bosons are, in the notation
of \cite{LongHP}, two neutral CP-even Higgs bosons, $H_1$ and
$H_2$, a pair of charged ones, $H^\pm_5$, and a massless neutral
CP-odd Higgs boson, $A_2$. The last one would receive its mass
through radiative corrections. As for the heavy Higgs spectrum, it
is comprised by two neutral CP-even scalar bosons, $H_3$ and
$H^\prime_4$, a neutral CP-odd one, $A_1$, and a pair of charged
ones, $H^\pm_6$. As pointed out in Ref. \cite{LongHP}, the neutral
CP-even scalar boson $H_2$ coincides with the SM one provided that
$\lambda_4=\lambda_5$. For the purpose of this work, it is enough
to consider this scenario. For the sake of clarity, we will only
present the expressions which relates the gauge states to the mass
eigenstates of the light sector. The real part of the $\rho^0$ and
$\eta^0$ neutral components of $\Phi_1$ and $\Phi_2$ define the
CP-even states $H_1$ and $H_2$ via the following rotation
\begin{eqnarray}
H_1&=&c_\beta \eta^0_r-s_\beta \rho^0_r, \\
H_2&=&s_\beta \eta^0_r+c_\beta \rho^0_r,
\end{eqnarray}
where $\beta$ is defined by $\tan\beta=(v_2/v_1)$,
$c_\beta=\cos\beta$, $s_\beta=\sin\beta$, and the subscript $r$
denotes the real part of the field. In the charged sector, the
$\rho^+$ and $\eta^+$ components of $\Phi_1$ and $\Phi_2$ define
the charged $H^+_5$ Higgs boson and the pseudo-Goldstone boson
associated with the $W$ gauge boson $G^+_W$:
\begin{eqnarray}
H^+_5&=&c_\beta \eta^++s_\beta \rho^+, \\
G^+_W&=&-s_\beta \eta^++c_\beta \rho^+.
\end{eqnarray}
Finally, the imaginary part of $\rho^0$ and $\eta^0$ define the
pseudo-Goldstone boson associated with the $Z$ gauge boson $G_Z$
and the massless CP-odd scalar boson $A_2$.

\begin{figure}[!hbt]
\centering
\includegraphics[height=4in]{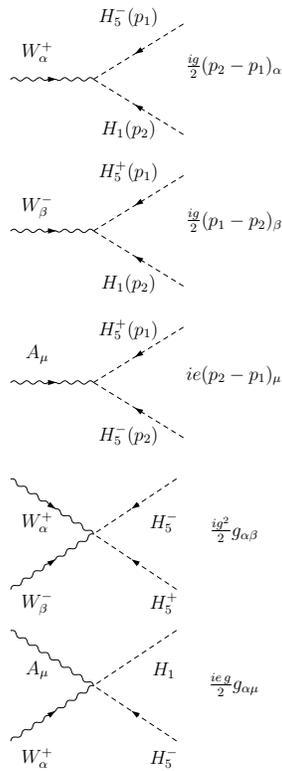}
\caption{\label{FeynRules-scalar}Feynman rules for the vertices
which contribute to the on-shell $WW\gamma$ vertex in the scalar
sector of the 331 model with right-handed neutrinos. The arrows
represent the flow of the 4-momenta. The coupling of $H_2$ to the
$W$ gauge boson is SM-like.}
\end{figure}

Once the light Higgs mass eigenstates are defined, from the
Higgs-kinetic term it is straightforward to obtain those couplings
involving the $W$ gauge boson. In order to analyze the behavior of
the Higgs sector at the Fermi scale in the scenario with
$\lambda_4=\lambda_5$, we will present the full Lagrangian
involving the couplings of the $W$ boson to the neutral and
charged Higgs bosons. It can be written as
\begin{equation}
\mathcal{L}=\left(m^2_W+g\,m_W\,H_2+\frac{g^2}{4}\left(H^2_1+H^2_2+2H^-_5H^+_5\right)\right)W^-_\mu
W^{+\mu}.
\end{equation}
There is a similar expression involving the $Z$ boson. It is also
interesting to note that there is no trilinear self-coupling of
the $H_1$ Higgs boson. From this Lagrangian it is evident that the
couplings of $H_2$ to the SM gauge bosons are SM-like, which means
that it should be identified with the SM Higgs boson. So its
contribution to the $WW\gamma$ vertex should be considered as a
part of the SM \cite{Bardeen} rather than a new physics effect. In
fact, the only contribution which can be considered as a new
physics effect is that induced by the $H^-_5H^+_5WW$ vertex. The
model also induces the trilinear $W^\pm H^\mp_5 H_1$ and quartic
$\gamma W^\pm H_1H^\mp_5$ vertices, which  also can contribute to
the $WW\gamma$ coupling. The corresponding Lagrangian for these
terms can be written as
\begin{equation}
\mathcal{L}=\frac{ig}{2}\left(W^+_\mu\left(H_1\partial^\mu
H^-_5-H^-_5\partial^\mu H_1\right)-W^-_\mu\left(H_1\partial^\mu
H^+_5-H^+_5\partial^\mu
H_1\right)\right)+\frac{eg}{2}H_1A_\mu\left(W^{+\mu}H^-_5+W^{-\mu}H^+_5\right).
\end{equation}
There is no analogous Lagrangian for $H_2$, which is in agreement
with the fact that $H_2$ plays the role of the SM Higgs boson. It
is worth noting that all of the couplings which contribute to the
$WW\gamma$ vertex are determined entirely by the coupling constant
$g$, in contrast with the case of the most general THDM potential,
which involves mixing angles. This fact will simplify considerably
the analysis of the $\Delta \kappa$ and $\Delta Q$ form factors as
they will depend only on the ${H_1}$ and ${H^+_5}$ masses, which
resembles the situation arising in the gauge sector, where the
form factors depend only on the bilepton masses. In particular,
since both $H_1$ and $H^+_5$ receive their masses at the Fermi
scale, it is also reasonable to analyze the scenario in which they
are degenerate. From the above Lagrangians it is straightforward
to obtain the Feynman rules necessary for our calculation. For the
sake of completeness they are shown in Fig.
\ref{FeynRules-scalar}.

\section{Static quantities of the $W$ boson}
\label{calculation}

In the usual notation, the most general CP-even on-shell
$WW\gamma$ vertex can be written as \cite{Bardeen}

\begin{equation}
\Gamma^{\mu \alpha \beta}=i\,e\left\{A\left[2\,p^\mu g^{\alpha
\beta}+4\,\left(Q^\beta\,g^{\mu \alpha}-Q^\alpha\,g^{\mu
\beta}\right)\right] +\,\Delta \kappa \left(Q^\beta\,g^{\mu
\alpha}-Q^\alpha\,g^{\mu \beta}\right)+\frac{4 \,\Delta Q
}{m_W^2}p^\mu \,Q^\alpha\,Q^\beta\right\},\label{WWg}
\end{equation}

\noindent where the momenta of the particles are denoted as
follows. $(p-Q)_\alpha$ and $(P+Q)_\beta$ are the momenta of the
incoming and outgoing $W$ boson, respectively, and $2Q_\mu$ is
that of the photon. We have dropped the CP-odd terms since they do
not arise in the 331 model with right-handed neutrinos. This class
of terms can be generated for instance in models in which the $W$
boson couples to both left- and right-handed fermions
simultaneously \cite{Burgess}. Both $\Delta \kappa$ and $\Delta Q$
vanish at the tree level in the SM, and the one-loop corrections
from the fermion, gauge, and scalar sectors are all of the order
of $\alpha/\pi$ \cite{Bardeen}. These form factors define the
magnetic dipole moment ($\mu_W$) and the electric quadrupole
moment ($Q_W$) of the $W$ boson, which are given by

\begin{equation}
\mu_W=\frac{e}{2\,m_W}\,\left(2+\Delta \kappa \right),
\end{equation}

\begin{equation}
Q_W=-\frac{e}{m_W^2}\,\left(1+\Delta \kappa +\Delta Q\right).
\end{equation}

It is interesting to note that the gauge invariant form
(\ref{WWg}) is obtained only after adding up the full
contributions of a particular sector of any specific model. Gauge
invariance along with the cancellation of ultraviolet divergences
are thus a test to check the correctness of the result.

In this work we are interested in the contribution to $\Delta
\kappa$ and $\Delta Q$ from the 331 model with right-handed
neutrinos. As already explained, the exotic quarks do not
contribute to the $WW\gamma$ vertex, whereas the extra neutral
boson $Z^\prime$ contribution arises from $Z$-$Z^\prime$ mixing
and it is expected to be negligibly small. The only contributions
to $\Delta \kappa$ and $\Delta Q$ arise from the gauge and scalar
sectors. In the former, the static properties of the $W$ boson
receive contributions from both the neutral and singly-charged
bileptons. As far as the scalar sector is concerned, it
contributes via the neutral and singly charged Higgs bosons.
Before presenting the results for these contributions, we would
like to comment briefly on our calculation scheme, which has been
already discussed in Refs. \cite{Stuart,Tavares}.

Rather than solving the loop integrals by the Feynman parameters
technique, one alternative is to use the Passarino-Veltman method
\cite{Passarino} to reduce the tensor integrals down to scalar
functions. However, this scheme cannot be applied together with
the kinematic condition $Q^2=0$ since it requires the inversion of
a matrix whose Gram determinant is directly proportional to $Q^2$.
Nevertheless, the Passarino-Veltman reduction scheme can be safely
applied for $Q^2\ne 0$, and the limit $Q^2\to 0$ can be taken at
the end of the calculation, which usually requires the application
of l' H\^opital rule: $\lim_{Q^2\to 0} f(Q^2)/Q^2=f^\prime(0)$.
This means that the result is given in terms of scalar functions
and its derivatives. It was shown in Ref. \cite{Stuart} that any
$N$-point scalar function and its derivatives with respect to any
of its arguments can be expressed in terms of a set of
$(N-1)$-point scalar functions when the kinematic Gram determinant
vanishes. It follows that one can express the three-point scalar
function $C_0$ appearing in the calculation and its derivative
with respect to $Q^2$ in terms of two-point scalar functions
$B_0$. The explicit reduction was presented in Ref. \cite{Stuart}.
It is then straightforward to obtain the limit $Q^2\to 0$. The
advantages of this method are twofold: it can be implemented in a
computer program \cite{Mertig}, which avoids the risk of any
error, and the two-point scalar functions can be readily solved
analytically or numerically \cite{FF}. This calculation scheme is
suited to solve loop diagrams carrying vector bosons, which may
give rise to some cumbersome tensor integrals.

\subsection{Gauge boson contribution}

\begin{figure}
 \centering
\includegraphics[width=3in]{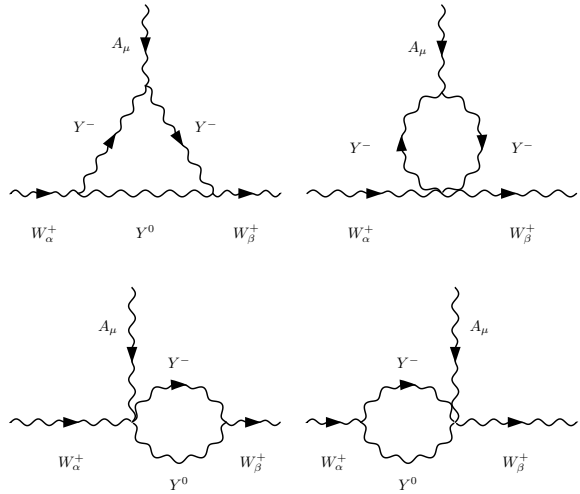}
\caption{\label{Feyn-Diag}Feynman diagrams, in the unitary gauge,
for the contribution to the on-shell $WW\gamma$ vertex from the
gauge sector of the 331 model with right-handed neutrinos.}
\end{figure}

We turn now to the contributions to $\Delta Q$ and $\Delta \kappa$
from the Feynman diagrams of Fig. \ref{Feyn-Diag}. The amplitudes
for these diagrams can be written down with the help of the
Feynman rules shown in Fig. \ref{FeynRules}. After applying the
calculation scheme described earlier and taking into account Eq.
(\ref{WWg}) we are left with

\begin{equation}
\label{DQ_eq} \Delta Q^Y=a\,\left(\frac{12\,\xi\,\eta-\chi^2
}{2\,\xi\,\eta}\right)\left(f_0(\xi,\eta)+f_1(\xi,\eta)\,\log\left(\frac{\eta}{\xi}\right)+
f_2(\xi,\eta)\,{\rm
arccot}\left(\frac{\xi+\eta-1}{\chi}\right)\right),
\end{equation}
and
\begin{equation}
\label{Dk_eq} \Delta \kappa^Y=\frac{a}{4\,
\xi^2\,\eta}\left(g_0(\xi,\eta)+g_1(\xi,\eta)\,\log\left(\frac{\eta}{\xi}\right)+
g_2(\xi,\eta)\,{\rm
arccot}\left(\frac{\xi+\eta-1}{\chi}\right)\right),
\end{equation}
where we have introduced the definitions $a=g^2/(96\,\pi^2)$,
$\xi=m_{Y^\pm}^2/m_W^2$, $\eta=m_{Y^0}^2/m_W^2$, and
$\chi^2=4\,\xi\,\eta-(\xi+\eta-1)^2$. The $f_i$ and $g_i$
functions read
\begin{align}
f_0(\xi,\eta)&=-\frac{2}{3}- 2\,\left(\xi - \eta \right)^2+ 3\,\xi
-\eta,
\end{align}
\begin{align}
f_1(\xi,\eta)&=\left( {\left( \eta - \xi  \right) }^2 - 2\,\xi
\right) \,\left( \eta  - \xi  \right) - \xi,
\end{align}
\begin{align}
f_2(\xi,\eta)&=-\frac{2}{\chi}\left(\left({\xi-\eta
}\right)^4-{\eta }^3 - \xi  - \eta \,\left( 1 + \eta  \right)
\,\xi  +\left( 3 + 5\,\eta  \right) \,{\xi }^2 - 3\,{\xi }^3
\right),
\end{align}
\begin{align} g_0(\xi,\eta)&=9\,{\eta }^3 + 6\,{\eta }^4 +
{\left(\xi -1 \right) }^2\,\left( 1 + \xi \,\left( 7 + 16\,\xi
\right)  \right) -{\eta }^2\,\left( 35 + \xi \,\left( 59 + 98\,\xi
\right) \right) +\eta \,\left( 19 + \xi \,\left( 70 + \xi \,\left(
3 + 4\,\xi \right) \right)\right),
\end{align}
\begin{align}
g_1(\xi,\eta)&=-3\,{\eta }^5 + 3\,{\eta }^4\,\left(\xi-1 \right)
-2\,\eta \,\left( \xi-1  \right) \,\xi \,\left( 3 + \xi \right) \,
\left(3\,\xi-1  \right)  +2\,{\left(\xi-1 \right) }^3\,\xi
\,\left( 1 + 4\,\xi  \right)  +{\eta }^3\,\left( 15 + \xi \,\left(
32 + 49\,\xi  \right)  \right)\nonumber\\&-3\,{\eta }^2\,\left( 3
+ \xi \,\left( 9 + \xi \,\left( 3 + 17\,\xi  \right)  \right)
\right),
\end{align}
\begin{align}
g_2(\xi,\eta)&=2\,\Big( 3\,{\eta }^6 - 6\,{\eta }^5\,\xi +
2\,{\left(\xi-1  \right) }^4\,\xi \,\left( 1 + 4\,\xi  \right) -
2\,\eta \,{\left(\xi-1  \right) }^2\,\xi \,\left( 2 + \xi \right)
\,\left(7\,\xi -1 \right)  - 2\,{\eta }^4\,\left( 9 + \xi \,\left(
19 + 23\,\xi  \right)  \right)\nonumber\\  &+ 2\,{\eta }^3\,\left(
12 + \xi \, \left( 31 + 45\,\xi  + 50\,{\xi }^2 \right)  \right) -
3\,{\eta }^2\,\left( 3 + \xi \, \left( 8 + \xi \,\left( 22 + \xi
\,\left(15\,\xi -16   \right) \right)  \right) \right)  \Big).
\end{align}

One interesting scenario is that in which the bileptons are
degenerate, {\it i.e.} $m_Y^\pm=m_Y^0=m_Y$, which is actually a
good assumption for $m_{Y^0}$ much larger than $m_W$ because of
the mass splitting (\ref{splitting}). In this scenario we obtain

\begin{equation}
\label{DQ_deg} \Delta Q^Y=\frac{a}{\zeta}\,\left( 1 + 4\,\zeta
\,\left(3\,\zeta-1 \right) \right)\left(1-\frac{1
}{3\,{\zeta}}-\frac{2\,\zeta-1}{{\sqrt{4\,\zeta-1 }}}{\rm
arccot}\left(\frac{ 2\,\zeta -1}{{\sqrt{4\,\zeta-1
}}}\right)\right),
\end{equation}
and
\begin{equation}
\label{Dk_deg} \Delta \kappa^Y=\frac{a}{2\,\zeta^2}\left(12 +
\zeta \, \left( 19 - 36\,\zeta \,\left( 1 + \zeta  \right)
\right) +\frac{1}{2\, {\zeta }}+\frac{\left(6\,\zeta-1  \right)
\,\left( \zeta \,\left(12\,\zeta \,\left( 1 + \zeta  \right)-7
\right)-2\right) }{{\sqrt{4\,\zeta-1 }}}{\rm arccot}\left(\frac{
2\,\zeta -1}{{\sqrt{4\,\zeta-1 }}}\right)\right),
\end{equation}
with $\zeta=m_Y^2/m_W^2$. It can be shown that in the heavy
bilepton limit both (\ref{DQ_deg}) and (\ref{Dk_deg}) behave as
$m_W^2/m_Y^2$ at the leading order in $m_Y$. It is then evident
that both form factors are insensitive to nondecoupling effects of
heavy bileptons.

\subsection{Scalar contribution}

\begin{figure}
 \centering
\includegraphics[width=3in]{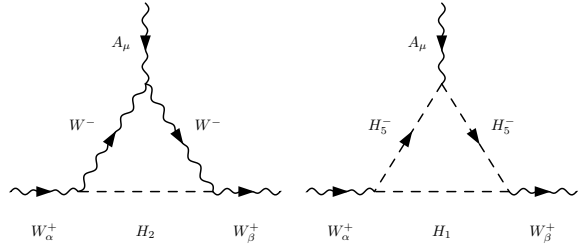}
\caption{\label{Feyn-Diag-scalar}Feynman diagrams for the scalar
contribution to the on-shell $WW\gamma$ vertex in the 331 model
with right-handed neutrinos. Although the $H_1$ and $H^\pm_5$
scalar bosons also induce three two-point diagrams, they do not
contribute to $\Delta \kappa$ or $\Delta Q$. In the scenario
described in the text, $H_2$ coincides with the SM Higgs boson, so
the left-hand triangle contribution belongs to the SM and will not
be considered a new physics effect.}
\end{figure}

In the scenario discussed earlier, the $W$ electromagnetic form
factors are induced by the charged scalar $H_5^\pm$ and the
neutral scalar bosons $H_1$ and $H_2$, which give rise to the
Feynman diagrams shown in Fig. \ref{Feyn-Diag-scalar}. The Feynman
rules necessary for this calculation are shown in Fig.
\ref{FeynRules-scalar}. We would like to mention that the neutral
$H_1$ and the charged scalar $H_5^\pm$ bosons induce three
additional two-point diagrams, but they are not shown in Fig.
\ref{Feyn-Diag-scalar} as give no contribution to the
electromagnetic form factors. Since the neutral Higgs boson $H_2$
coincides with the SM one, the contribution from the triangle
diagram of the left-hand side of Fig. \ref{Feyn-Diag-scalar} is in
fact a SM effect rather than a new physics effect. The result for
this contribution was obtained long ago \cite{Bardeen,Argyres}. As
for the Feynman diagram of the right-hand side, it yields a
contribution similar to that arising in the THDM, as can be
inferred from the Feynman rules given in Fig.
\ref{FeynRules-scalar}. Although this contribution was already
obtained in terms of Feynman-parameter integrals \cite{THDM}, we
would like to present an alternative result in terms of elementary
functions. The calculation scheme described above yields

\begin{equation}
\label{DQ_Higgs} \Delta Q^H=a\,\left({\tilde
f}_0(\lambda,{\lambda_{+}})+{\tilde f}_1(\lambda,
{\lambda_{+}})\,\log\left(\frac{{\lambda_{+}}}{\lambda}\right)+
{\tilde f}_2(\lambda,{\lambda_{+}})\,{\rm
arccot}\left(\frac{\lambda+{\lambda_{+}}-1}{{\widetilde
\chi}}\right)\right),
\end{equation}
and
\begin{equation}
\label{Dk_Higgs} \Delta \kappa^H=a\,\left({\tilde
g}_0(\lambda,{\lambda_{+}})+{\tilde g}_1(\lambda,
{\lambda_{+}})\,\log\left(\frac{{\lambda_{+}}}{\lambda}\right)+
{\tilde g}_2(\lambda,{\lambda_{+}})\,{\rm
arccot}\left(\frac{\lambda+{\lambda_{+}}-1}{{\widetilde
\chi}}\right)\right),
\end{equation}
with
\begin{align}
{\tilde f}_0(\lambda,{\lambda_{+}})&=\frac{2}{3} + {\lambda_{+}}
\,\left(2\,{\lambda_{+}}-3 \right) + \lambda - 4\,{\lambda_{+}}
\,\lambda +
  2\,{\lambda }^2,
\end{align}
\begin{align}
{\tilde f}_1(\lambda,{\lambda_{+}})&=-{\lambda_{+}}\,
\left({\lambda_{+}}-1 \right)^2+ {\lambda_{+}} \,\lambda
\,\left(3\,{\lambda_{+}}  -2 \right)  - 3\,{\lambda_{+}}
\,{\lambda }^2 + {\lambda }^3,
\end{align}
\begin{align}
{\tilde f}_2(\lambda,{\lambda_{+}})&=\frac{2}{{{\widetilde \chi}
}}\,\left( {\lambda_{+}} \,\left({\lambda_{+}}-1 \right)^3 -
{\lambda_{+}}\,\lambda\,\left({\lambda_{+}}-1 \right)
\,\left(4\,{\lambda_{+}}-1 \right)
 + {\lambda_{+}} \,\left(6\,{\lambda_{+}}-1  \right) \,{\lambda }^2 - \left( 1 +
4\,{\lambda_{+}} \right) \,{\lambda }^3 + {\lambda }^4 \right),
\end{align}
\begin{align}
{\tilde g}_0(\lambda,{\lambda_{+}})&=2 - {\lambda_{+}}  +
4\,\lambda,
\end{align}
\begin{align}
{\tilde g}_1(\lambda,{\lambda_{+}})&=\frac{1}{2}\left(1 +
\left({\lambda_{+}}-2\right) \,{\lambda_{+}} + \lambda -
5\,{\lambda_{+}} \,\lambda + 4\,{\lambda }^2\right),
\end{align}
\begin{align}
{\tilde g}_2(\lambda,{\lambda_{+}})&=- \frac{1}{{\widetilde
\chi}}\left(\left({\lambda_{+}}-1 \right)^3 -
6\,\left({\lambda_{+}}-1  \right) \,{\lambda_{+}} \,\lambda +
3\,\left( 1 + 3\,{\lambda_{+}} \right) \,{\lambda }^2 -
4\,{\lambda }^3 \right).
\end{align}
In these equations, $\lambda$, ${\lambda_{+}}$ and ${\widetilde
\chi}$ are given as follows.
 $\lambda=m_{H_1}^2/m_W^2$,
${\lambda_{+}}=m_{H^\pm_5}^2/m_W^2$, and ${\widetilde
\chi}^2=4\,\lambda\,{\lambda_{+}}-(\lambda+{\lambda_{+}}-1)^2$.

When both the neutral and the charged Higgs bosons are mass
degenerate ($\lambda_+=\lambda=\tilde\zeta$) Eqs. (\ref{DQ_Higgs})
and (\ref{Dk_Higgs}) yield

\begin{equation}
\label{DQ_Higgs_deg} \Delta Q^H=\frac{2\,a }{3}\left( 1 -
3\,{\tilde\zeta} + \frac{3\,{\tilde\zeta}
\,\left(2\,{\tilde\zeta}-1 \right) \, }{{\sqrt{4\,{\lambda}-1
}}}{\rm arccot}\left(\frac{ 2\,{\lambda} -1}{{\sqrt{4\,{\lambda}-1
}}}\right)\right),
\end{equation}
and
\begin{equation}
\label{Dk_Higgs_deg} \Delta \kappa^H=a\left(2 + 3\,{\tilde\zeta} +
\frac{\left( 1 - 3\,{\tilde\zeta}\,\left(1  +
2\,{\tilde\zeta}\right) \right) \, }{{\sqrt{ 4\,{\lambda}-1
}}}{\rm arccot}\left(\frac{ 2\,{\lambda} -1}{{\sqrt{4\,{\lambda}-1
}}}\right)\right).
\end{equation}

We have numerically evaluated the above results and found
agreement with those obtained in Ref. \cite{THDM}.

\section{Numerical evaluation}
\label{discussion}

We will now analyze the behavior of the form factors for some
range of values of the bilepton and the Higgs boson masses. These
are the only free parameters which enter into $\Delta Q$ and
$\Delta \kappa$. We will analyze separately each contribution.

\subsection{Gauge boson contribution}

To begin with, it is worth analyzing the current bounds on the
bilepton masses from both theoretical and experimental grounds.
First of all, it is interesting to note that the matching of the
gauge couplings constants at the ${SU_L(3)}\times {U_X(1)}$
breaking leads to $4\,\sin \theta_W \le 1$ in the minimal 331
model \cite{Pleitez1}, from which an upper bound on the bilepton
masses can be derived, namely $m_Y \lesssim 1$ TeV. Therefore,
this model would be confirmed or ruled out by collider experiments
in a near future. The version with right-handed neutrinos requires
however that $4\,\sin \theta_W \le 3$, which yields no useful
constraint on $m_Y$. As already mentioned, because of the symmetry
breaking hierarchy, the splitting
$\left|m_{Y\pm}^2-m_{Y^0}^2\right|$ is bounded by the $W$ boson
mass. Therefore $m_{Y^0}$ and $m_{Y^\pm}$ are not arbitrary at
all. One cannot, for instance, make large $m_{Y\pm}$ while keeping
fixed $m_{Y^0}$ or viceversa. In fact, when $m_{Y^0}\gg m_{W}$,
the charged and neutral bileptons would become degenerate. As far
as the lower bounds on the bilepton masses are concerned, in Ref.
\cite{Long} it was argued that the data from neutrino neutral
current elastic scattering give a lower bound on the mass of the
new neutral gauge boson $m_{Z_2}$ in the range of 300 GeV, which
along with the symmetry-breaking hierarchy yield $m_{Y^\pm} \sim
m_{Y_0} \sim 0.72\, m_{Z_2}\ge 220$ GeV. A similar bound was
obtained in Ref. \cite{Long-Inami} from the observed limit on the
``wrong'' muon decay $R=\Gamma(\mu^-\to e^- \nu_{e}
\bar{\nu}_\mu)/\Gamma(\mu^-\to e^- \bar{\nu}_{e} \nu_\mu)\le
1.2\%$, which leads to $m_{Y^\pm}\ge 230\pm 17$ GeV at 90\% C.L.
These lower bounds on $m_Y^{\pm}$ are in agreement with that
obtained from the latest BNL measurement on the muon anomaly
\cite{Long,Bounds}.

According to the above discussion, we deem it interesting to
evaluate the form factors in the range 100 GeV $\le m_{Y^0}\le$
1000 GeV, which will be useful to illustrate their behavior and
get an idea about their size. At this point it is important to
mention that to cross-check our results, the form factors were
obtained independently by the Feynman parameters method. The
integrals were evaluated numerically and the result was compared
with the one obtained by the Passarino-Veltman method. A perfect
agreement was observed. We refrain from presenting the results in
terms of parametric integrals since the closed expressions
(\ref{DQ_eq}) and (\ref{Dk_eq}) can be handled more easily.

\begin{figure}
\centering
\includegraphics[width=2.5in]{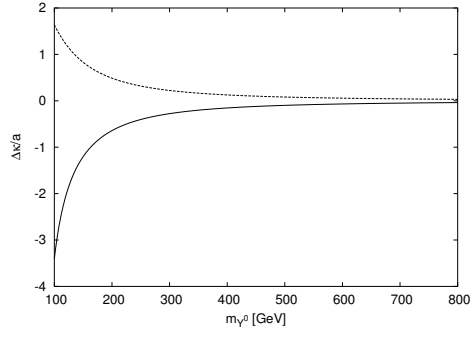}
\caption{\label{DK_Y0}Gauge boson contribution to $\Delta \kappa$
in the 331 model with right-handed neutrinos as a function of the
mass of the neutral bilepton when the charged bilepton mass is
maximal (solid line) and minimal (dashed line). According to the
mass splitting, the extremal values are given by
$m_{Y^\pm}^2=m_{Y^0}^2\mp m_W^2$. The form factor is restricted to
lie in the area enclosed by the lines.}
\end{figure}

\begin{figure}
\centering
\includegraphics[width=2.5in]{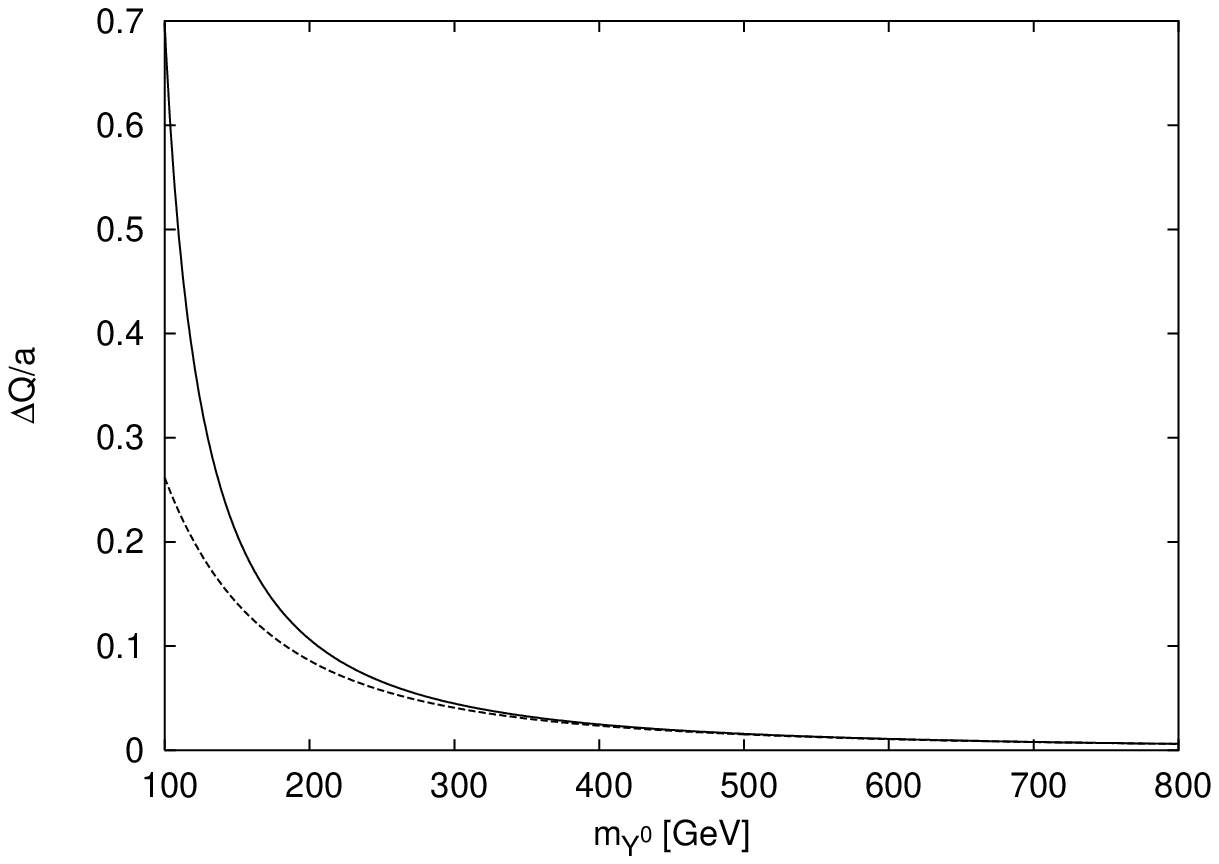}
\caption{\label{DQ_Y0}The same as in Fig. \ref{DK_Y0} for the
$\Delta Q$ form factor.}
\end{figure}

\begin{figure}
\centering
\includegraphics[width=2.5in]{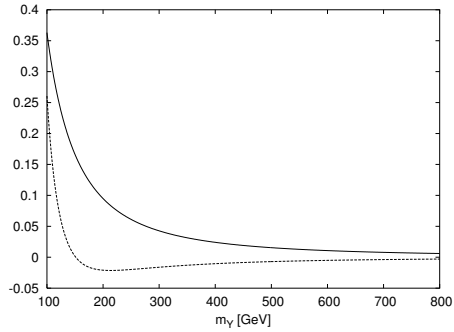}
\caption{\label{DQDK_Y0_Deg}Bilepton contribution to the $\Delta
\kappa$ (solid line) and $\Delta Q$ (dashed line) form factors, in
units of $a$, in the 331 model with right-handed neutrinos when
the bileptons are degenerate and have a mass $m_Y$.}
\end{figure}

The $\Delta \kappa^Y$ and $\Delta Q^Y$ form factors are shown in
Fig. \ref{DK_Y0} and Fig. \ref{DQ_Y0} as a function of the neutral
bilepton mass. There are two curves in each plot, which correspond
to the extremal values of $m_{Y^\pm}$, namely
$m_{Y^\pm}^2=m_{Y^0}^2- m_W^2$ and $m_{Y^\pm}^2=m_{Y^0}^2+ m_W^2$.
The form factors are restricted to lie in the area surrounded by
the two extremal lines. In Fig. \ref{DK_Y0} it is clear that the
bileptons can give a negative or positive contribution to $\Delta
\kappa^Y$, which depends on which bilepton is the heaviest. Also,
we can observe that $\Delta \kappa^Y$ is sensitive to the value of
the splitting and has a larger size for nondegenerate bileptons
than for degenerate bileptons. The $\Delta \kappa^Y$ form factor
in the latter scenario is displayed in Fig. \ref{DQDK_Y0_Deg}. In
this plot we can observe that, when one of the bilepton masses is
close to $m_W$ and the splitting is maximal, $\Delta \kappa^Y$ can
have a size of about one order of magnitude above than the one
obtained when there is degeneracy of the bilepton masses. On the
contrary, $\Delta Q^Y$ is less sensitive to the mass splitting and
the extremal values of $m_{Y^\pm}$ yield values of the same order
of magnitude than the one observed in the degenerate case, which
is also shown in Fig. \ref{DQDK_Y0_Deg}. From these plots we can
conclude that the size of the contribution to the form factors
from the gauge sector of the 331 model with right-handed neutrinos
is about of the same order of magnitude than the one obtained in
the case of the bilepton contribution in the minimal 331 model and
in the case of the contributions of other SM extensions. The
larger absolute values are obtained for lighter bileptons and when
the charged bilepton mass reaches its maximal allowed value. It is
interesting to note that all weakly coupled theories studied up to
now give a contribution to the $W$ form factors of similar size
\cite{THDM,Tavares,WWg-NP}.

In Fig. \ref{DQDK_Y0_Deg}, we can clearly see that both $\Delta
\kappa^Y$ and $\Delta Q^Y$ are insensitive to heavy physics
effects and approach to zero very quickly as the bilepton masses
increase. The only scenario which may give rise to nondecoupling
effects is that in which one bilepton mass is kept fixed while the
other is made very large, which of course is forbidden by the mass
splitting constraint (\ref{splitting}). In Ref. \cite{Tavares} we
already discussed a similar situation arising in the minimal 331
model, with a doubly charged bilepton playing the role of the
neutral one. This case also resembles the one discussed in Ref.
\cite{Li} for a scalar doublet which acquires mass from a bare
parameter. The reason why there is no decoupling effects is not
surprising since a large bilepton mass implies a large VEV which
is heavier than the electroweak scale. On the contrary, the
splitting between the bilepton masses arises from VEVs which are
of the size of the electroweak scale. This is to be contrasted
with the case of a fermion pair accommodated in a ${SU_L(2)}$
doublet, which are known to give rise to nondecoupling effects.
Since the fermions acquire their masses from Yukawa couplings, a
large fermion mass implies a large coupling, whereas a heavy
bilepton mass implies a large VEV instead of a large coupling. The
former scenario is the one which is known to break down the
decoupling theorem \cite{Appelquist}. Very interestingly, even in
those scenarios in which $\Delta \kappa^Y$ is sensitive to heavy
physics effects, $\Delta Q^Y$ is not \cite{Inami}.  The decoupling
theorem establishes that only those terms arising from
renormalizable operators may be sensitive to nondecoupling
effects, whereas those terms induced by nonrenormalizable
operators are suppressed by inverse powers of the heavy mass
\cite{Appelquist}. Thus $\Delta Q^Y$ always decouples when one
particle circulating in the loop is made large since it is
generated by a nonrenormalizable dimension-six operator, but
$\Delta \kappa^Y$ may be sensitive to nondecoupling effects as it
is induced by a dimension-four operator.

\subsection{Scalar contribution}

\begin{figure}
\centering
\includegraphics[width=2.5in]{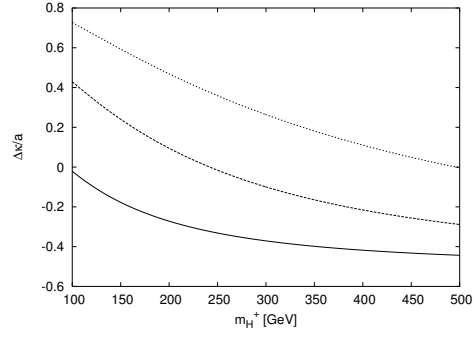}
\caption{\label{DK_scalar}Scalar contribution to $\Delta \kappa$
in the 331 model with right-handed neutrinos as a function of
$m_{H^\pm_5}$ for different values of the mass of the neutral
Higgs boson $m_{H_1}$: 115 GeV (solid line), 250 GeV (dashed line)
and  500 GeV (dahed-dotted line) .}
\end{figure}

\begin{figure}
\centering
\includegraphics[width=2.5in]{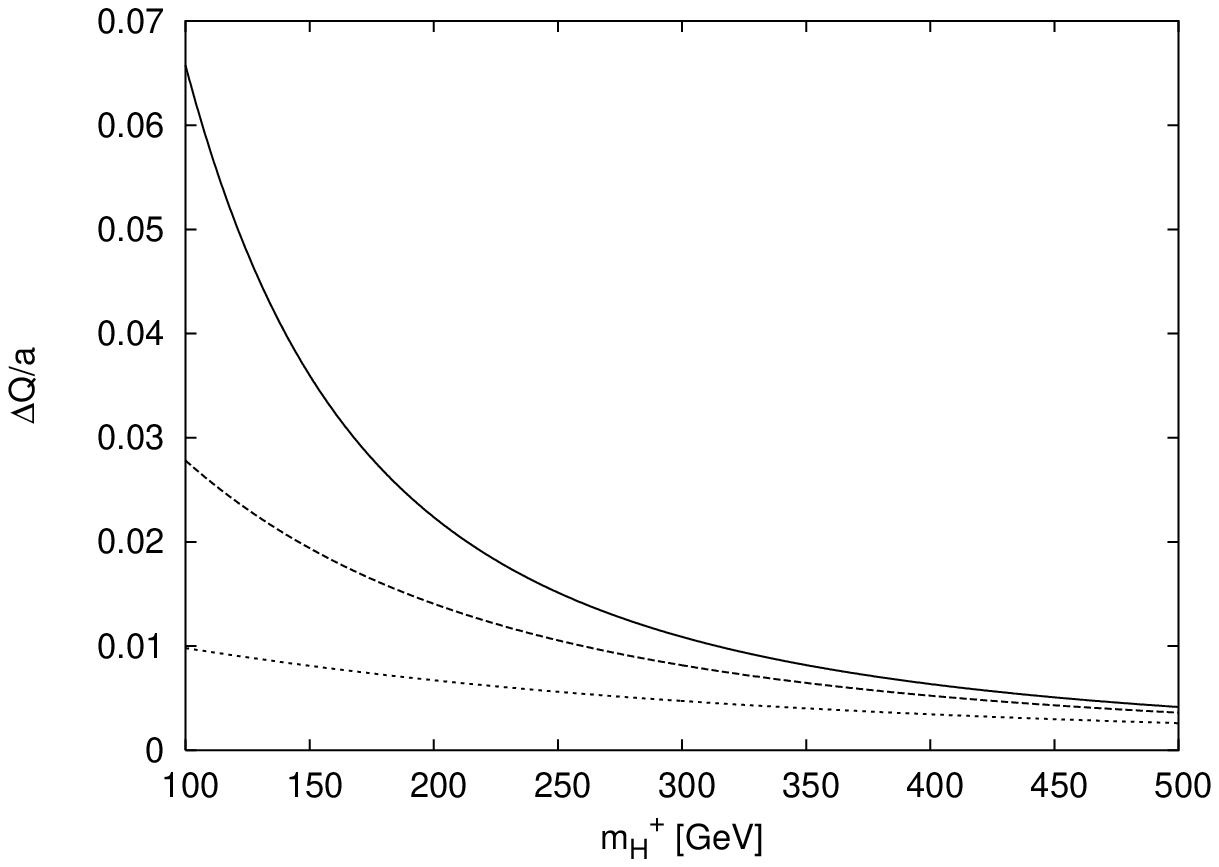}
\caption{\label{DQ_scalar}The same as in Fig. \ref{DK_Y0} for the
$\Delta Q$ form factor.}
\end{figure}

\begin{figure}
\centering
\includegraphics[width=2.5in]{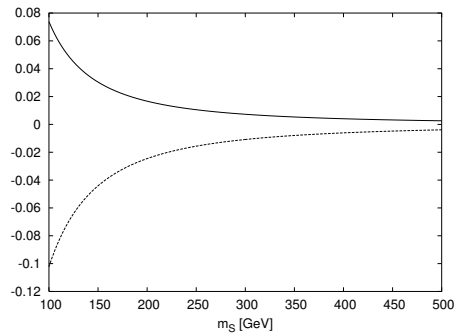}
\caption{\label{DQDK_scalar_Deg}Scalar contribution to the $\Delta
\kappa$ (solid line) and $\Delta Q$ (dashed line) form factors, in
units of $a$, in the 331 model with right-handed neutrinos when
$H_1$ and $H_5^\pm$ are degenerate and have a mass $m_S$.}
\end{figure}

The scalar contribution to the $\Delta \kappa$ and $\Delta Q$ form
factors is shown in Figs. \ref{DK_scalar} and \ref{DQ_scalar} as a
function of the charged Higgs boson mass and for different values
of the neutral Higgs boson mass. We would like to emphasize that
the values shown in those plots correspond to the contribution
from new physics only. From these Figs., we can observe that both
$\Delta \kappa^H$ and $\Delta Q^H$ decreases rapidly for
increasing $m_{H^\pm_5}$. In fact, the latter goes to zero quickly
as either $m_{H_1}$ or $m_{H^\pm_5}$ increase. Although $\Delta
\kappa^H$ seems to increase with increasing $m_{H_1}$ for a
relatively light $m_{H^\pm_5}$, it approaches the limiting value
$\Delta \kappa^H =a$ for very large $m_{H_1}$. We can also observe
that when the scalar boson masses are of the same size than those
of the bilepton gauge bosons, the contribution from the Higgs
sector is about one order of magnitude below than that of the
gauge sector. In fact, if the scalar boson masses are degenerate,
the respective contribution to $\Delta \kappa$ and $\Delta Q$ is
very small, as shown in Fig. \ref{DQDK_scalar_Deg}. As pointed out
in Ref. \cite{THDM}, this is a reflect from the fact that the
Higgs boson is not strongly interacting. Thus, for the Higgs
sector to give a large correction to the $W$ form factors, it
would be necessary to have the contributions from an unrealistic
number of Higgs bosons. Although we have restricted to a
particular form of the scalar potential, we can conclude that we
cannot expect large contributions from this sector even in the
most general case.

It is interesting to analyze the behavior of Eqs. (\ref{DQ_Higgs})
and (\ref{Dk_Higgs}) in the decoupling limit. It turns out that
$\Delta Q^H$ always vanishes no matter which one of $m_{H_1}$ or
$m_{H^\pm_5}$ is made large. On the other hand, $\Delta \kappa^H$
do may give rise to nondecoupling effects. If both $m_{H_1}$ and
$m_{H^\pm_5}$ become simultaneously large, $\Delta\kappa^H$
vanishes, but when $m_{H_1}$ becomes infinite and $m_{H^\pm_5}$
remains finite, it approaches the constant value
$\Delta\kappa^H=a$; when the situation is reversed, $\Delta
\kappa^H\to-a/2$. This is in accordance the previous discussion on
the decoupling properties of the $W$ form factors.

Finally, we would like to compare the size of the new
contributions with those of the SM, which is known to give the
following one-loop corrections to $\Delta \kappa$ and $\Delta Q$
\cite{Bardeen}: $\Delta \kappa^{\rm SM}_{\rm max}=30\,a$ and
$\Delta Q^{\rm SM}_{\rm max} =5\,a$. The contribution from the 331
model with right-handed neutrinos is thus only a few percent that
of the SM. From all the studies presented in the literature
\cite{WWg-NP}, it can be inferred that only those models in which
there are contributions from a large number of particles would
have the chance of giving large corrections to the $W$ form
factors.

\section{Final remarks}
\label{conclusions}

In this work we have calculated the static quantities of the $W$
boson in the framework of the 331 model with right-handed
neutrinos.  Apart from the usual SM contributions, there is new
contributions from the gauge and the scalar sectors. In the former
there is a new contribution induced by a singly charged $Y^\pm$
and a complex neutral gauge boson $Y^0$, called bileptons. In the
scalar sector there is the contribution from a singly charged
Higgs boson $H_5^\pm$ and two neutral scalar bosons $H_1$ and
$H_2$, but $H_2$ coincides with the SM Higgs boson and its
contribution should be identified with a SM effect rather than
with new physics. Although the model predicts three exotic quarks
and an extra neutral gauge boson $Z^\prime$, these particles give
no contribution to $\Delta Q$ and $\Delta \kappa$. It turns out
that the exotic quarks do not couple to the $W$ boson as they are
${ SU_L(2)}$ singlets, whereas $Z^\prime$ can only contribute
trough $Z-Z^\prime$ mixing and its contribution is expected to be
negligibly small. Analytical expressions were presented for both
nondegenerate and degenerate masses of the bileptons and the Higgs
bosons. The loop integrals were worked out by a modified version
of the Passarino-Veltman reduction scheme. To cross-check our
results, the form factors were obtained independently by the
Feynman parameter technique and the resulting integrals were
numerically evaluated and compared with the results obtained
through the Passarino-Veltman method. It was found that the new
contributions can be of the same order of magnitude as those
arising in other weakly coupled renormalizable theories. It is
interesting to note that the contribution from the scalar sector
is similar to that of a THDM. This means that the form factors
will not help to discriminate between different theories. Instead
the on-shell $WW\gamma$ vertex would be useful to test the
particular theory realized in nature with high precision once all
the free parameters of the theory are known. In the scenario in
which the non SM particles circulating in the loops (bileptons or
Higgs bosons) are degenerate, the form factors are smaller than in
the case in which they are nondegenerate. It was also found that
in the scenario in which the bilepton and scalar boson masses are
of the same order of magnitude, the gauge sector gives dominant
contribution to the $W$ form factors. The nondecoupling properties
of the $\Delta \kappa$ and $\Delta Q$ form factors were analyzed.
It was found that $\Delta Q$ is always of decoupling nature,
whereas $\Delta \kappa$ is sensitive to heavy Higgs bosons but
insensitive to heavy bileptons. In fact, the numerical analysis
shows that the contribution from a heavy bilepton with mass of the
order of 1 TeV is negligibly small.

\acknowledgments{Support from CONACYT and SNI  is acknowledged.
G.T.V also thanks partial support from SEP-PROMEP.}

\end{document}